\documentclass[preprint,showpacs,preprintnumbers,amsmath,amssymb]{revtex4}

\usepackage{graphicx}
\usepackage{dcolumn}
\usepackage{bm}

\newcommand{\qq}{\mbox{\boldmath$q$}} 
\newcommand{\pp}{\mbox{\boldmath$p$}}

\begin{document}

\title{Family of  additive entropy functions out of thermodynamic limit} 

\author{Alexander N. Gorban}
\email{gorban@icm.krasn.ru}
\affiliation{Institute of Computational Modeling RAS, 660036 Krasnoyarsk, Russia}

\author{Iliya V. Karlin}
\email{ikarlin@ifp.mat.ethz.ch}
\affiliation{
ETH Z\"{u}rich, Department of Materials, Institute of Polymers\\ 
ETH-Zentrum, Sonneggstr.\ 3, ML J 19, CH-8092 Z\"{u}rich, 
Switzerland
}

\date{\today}

\begin{abstract}
Starting with the additivity condition 
for Lyapunov functions of master equation,
we derive a one-parametric family of  entropy functions
which may be 
appropriate  for a description of certain  effects of finiteness of statistical systems,
in particular, distribution functions with long tails.
This one-parametric family  is different from Tsallis entropies, and
is essentially a convex combination of the Boltzmann-Gibbs-Shannon entropy
and the entropy function introduced by Burg. 
An example of how longer tails are described within the
present approach is worked out for the canonical ensemble.
In addition, we discuss a possible origin of a hidden statistical dependence,
and give explicit recipes how to construct corresponding generalizations of 
master equation.
\end{abstract}

\pacs{05.70.Ln, 05.20.Dd}
\maketitle

\section{Introduction}
\label{intro}

Past several years have witnessed a burst of interest in nonextensive
statistical mechanics, the topic which finds increasingly more applications particularly due
to the concept of Tsallis entropy \cite{Tsallis88}, \cite{Tpage}.  In this approach, one
postulates the following one-parametric family of concave functions,

\begin{equation} 
\label{Tsallis1} 
S_q=\frac{1-\sum_i p_i^q}{1-q}, 
\end{equation}
where $q>0$.  The family of Tsallis entropies (\ref{Tsallis1})
replaces the traditional Boltzmann-Gibbs-Shannon entropy,
$S_1$,
 
\begin{equation}
\label{BGS} S_1=\lim_{q\to1}S_q=-\sum_{i}p_i\ln p_i.  
\end{equation} 

One of the important achievements associated with Tsallis  entropy (\ref{Tsallis1})
is the fact that it provides an easy access -
through the  method of entropy maximization - to a rich set of
distribution functions, different from the traditional Gaussian distribution function, 
thereby giving a handy description of long (non-exponential) tails of probability distributions.
For each value of the parameter $q$, Tsallis entropy is a concave function
of the probability. 
The characteristic feature of Tsallis entropy is that it is nonextensive for $q\ne 1$,
that is, if the system is composed of two statistically independent subsystems,
Tsallis entropy of this system is not the sum of Tsallis entropies of the subsystems.
Since Tsallis entropy is postulated rather than derived, this point remains somewhat
open to discussions \cite{Vives02,KGG02}.

The goal of this paper is to present an argument on how long tails can be described in a usual,
extensive (more precisely, almost extensive) statistical mechanics, and to
give a theoretical derivation of a different, and in a certain sense unique,
one-parametric family of entropy functions which can model effects of finiteness. 

Our first remark is that real-world systems, to which statistical mechanics is applied, are finite,
and, although they do consist of a large number of subsystems, the natural logarithm of
this number (and since we address  questions related to entropies, one should observe the
magnitude of the logarithm, in the first place) is not that big after all, 
it is not larger than $100$ and is
often less than $20$.  Extensivity, in a true sense of this notion, theorems of
equivalence of the microcanonical and the canonical ensembles \cite{Ruelle}, and like,
is valid only in the thermodynamic limit where the system can be partitioned into an  arbitrary
large number of  noninteracting and statistically independent subsystems.
Namely, it is the number of such independent subsystems $\nu$ which do not interact, and
which are similar in all their observable properties to the larger system, that
plays the role of the parameter the values of which tells how close
is the system to the thermodynamic limit.

One is up to invoking that $\nu$  is finite (and  by doing so, one
restores to an argument about an incomplete extensivity) when one needs to cut off the
tails of distribution functions with divergent averages.
This fact is well known, for example,  in the case of the classical
Boltzmann equation: The maximum entropy
solution for the Boltzmann entropy does not exist (is not normalizable) if
the observables are the density, the average momentum,
the stress tensor and the heat flux
\cite{Koga54},
\cite{Kogan65},
\cite{Lewis67},
\cite{GK94b},
\cite{Levermore96}.
A regularization by the argument that the
magnitude of the microscopic velocity is restricted to
the value dictated by finiteness of the {\it total} energy \cite{Kogan65}
is an example of the incomplete extensivity argument. 

Thus, when the system is not strictly in the thermodynamic limit, details
of the interaction should gradually become more and more important, and prospects
of a {\it universal} description using a maximization of  an
{\it  interaction-independent} entropy functional
become less evident. Nevertheless, the very possibility of a sufficiently
reliable universal description in the sense just mentioned cannot  be  ruled out a priori.
For that reason, a search for non-classical entropies for a possible description
of nonextensive systems seems motivated.

The structure of this paper is as follows: In the next section \ref{Markov}, we
shall review, for the sake of completeness, the theory of Lyapunov functions
of master equation. In section \ref{entropy} we derive the family of the (almost)
additive entropies from the condition of additivity of the entropy function for
statistically independent systems. In section \ref{test} we demonstrate with
a simple example how  long tails are related to the effects of finiteness
in the present approach.

Sections \ref{entropy} and \ref{test} are the central point of our presentation.
In section \ref{ghost} we discuss a different scenario how the apparent statistical
dependence can occur when description of the system is incomplete,
and present a natural generalization of master equation for those cases in section
\ref{kinetics}. Finally, results are briefly discussed in section \ref{conclusion}.

\section{Lyapunov functions of master equation}\label{Markov}

We shall start our discussion  with a brief summary of the theory of Markov chains.
Our presentation essentially follows Ref.\  \cite{Gorban84}.
Let us consider a finite set of states $E_1,\dots,E_N$, and let us assume that the system
can be found only in those states. The probability of finding the system
at time $t\ge0$ is given by the $N$-component vector $\pp$ of probabilities $p_i(t)$,
such that $p_i\ge0$, and $\sum_{i=1}^N p_i=1$. Dynamic equation for $\pp$ of the form,
\begin{equation}
\label{Markov1}
\dot{p}_i=\sum_{k=1}^N q_{ik}p_{k},
\end{equation}
describes the time evolution of a Markov chain if and only if 
the matrix elements $q_{ik}$ satisfy $q_{ik}\ge0$ for $i\ne k$, and for every $k$,
\begin{equation}
\sum_{i=1}^N q_{ik}=0.
\end{equation}
For each Markov chain, a graph of transition is put into correspondence by
drawing an oriented link from the vertex $E_i$ to the vertex $E_k$ if $q_{ki}>0$.
Important class of Markov chains is characterized by directional connectivity.
The graph of transitions is called directionally connected if there is a path
from each vertex to any other vertex made up of the oriented links.
Then the following ergodic theorem is valid:
Let transition graph of the Markov chain be directionally connected. Then there
exists a positive stationary state $\pp^{\rm eq}$, $p^{\rm eq}_i>0$, and
for any initial condition $\pp(0)$, solution $\pp(t)$ to equation (\ref{Markov1})
tends to $\pp^*$ at $t\to\infty$.

Let the Markov chain satisfy to the ergodic theorem, and let the stationary
state $\pp^*$ is known. Let $h(x)$ be a convex and twice 
differentiable function of one variable
$x\in[0,\infty]$. Any function $h$ defines a convex Lyapunov function $H_h$ of the
Markov chain (\ref{Markov1}),

\begin{equation}
\label{Lyap1}
H_h(\pp)=\sum_{i=1}^N p_i^{\rm eq}h(p_i/p_i^{\rm eq}).
\end{equation}
Time derivative of the function $H_h$ (\ref{Lyap1}) due to dynamics (\ref{Markov1})
 is nonpositive:
\begin{equation}
\dot{H}_h=\sum_{i,j,i\ne j}q_{ij}p_j^{\rm eq}
[h(p_i/p_i^{\rm eq})-h(p_j/p_j^{\rm eq})
+ h'(p_i/p_i^{\rm eq})
((p_j/p_j^{\rm eq})-(p_i/p_i^{\rm eq}))]\le0,
\label{production1}
\end{equation}
where prime denotes derivative with respect to argument.
Equality sign is achieved only in the stationary state.

The stationary state $\pp^{\rm eq}$ is called the state of detail balance, if it satisfies

\begin{equation}
\label{DB}
q_{ik}p_k^{\rm eq}=q_{ki}p_i^{\rm eq}.
\end{equation}
Markov chain with detail balance is colloquially termed master equation.
In this case, the time derivative of the Lyapunov function becomes especially simple,
\begin{equation}
\dot{H}_h=-\frac{1}{2}\sum_{i,j,i\ne j}q_{ij}p_j^{\rm eq}
[  h'(p_i/p_i^{\rm eq})-
h'(p_j/p_j^{\rm eq})]
[(p_i/p_i^{\rm eq})-(p_j/p_j^{\rm eq})]\le0.
\label{production2}
\end{equation}
Discussion of the physical significance of the detail balance for the Markov chain
(master equation) is a well known textbook material. 

Since a convex linear combination of convex function is also a convex function,
the obvious construction which enables one to create
new Lyapunov functions from given representatives of the family (\ref{Lyap1}) is
this: Let $h_1,\dots,h_k$ be convex functions, and $\alpha_1,\dots,\alpha_k$
be nonnegative numbers, $\alpha_i\ge0$, satisfying $\sum_{m=1}^k\alpha_m=1$. Then,
\begin{equation}
\label{gen}
H_{\alpha_1 h_1\dots\alpha_k h_k}=\sum_{m=1}^k\sum_{i=1}^N
p_i^{\rm eq}\alpha_mh_m(p_i/p_i^{\rm eq}),
\end{equation}
is also the  Lyapunov function of Markov chain. It should be stressed that
the set (\ref{gen}) does not extends the family (\ref{Lyap1}) already specified.

Concluding this summary, we stress that, under physically
significant restrictions on the existence of the stationary state,
any Markov chain has a large class of Lyapunov functions of the form (\ref{Lyap1}),
each constructed from a convex function $h$ of one variable.
The additivity requirement makes it possible to drastically restrict the class of physically relevant
Lyapunov functions of Markov chains, which we do in the next section.

\section{Family of additive Lyapunov functions}
\label{entropy}

In order to derive the family of additive Lyapunov functions, 
let us consider two statistically independent systems described
by probability vectors $\pp$ and $\qq$, $p_i\ge 0$, $q_j\ge 0$,
$\sum_i p_i=1$, $\sum_j p_j=1$. respectively.
We shall consider first the case of the equipartition at the equilibrium, in order
to simplify notation. Thus, we  assume that 
equilibrium states of both the systems are equipartitions with
probability vectors $\pp^{\rm eq}$ and $\qq^{\rm eq}$, where
$p_i^{\rm eq}=1/P$, $q_i^{\rm eq}=1/Q$, and where $P$ and $Q$
are the numbers of states in the systems.

Since the systems are independent, 
the joint system is characterized by the joint probability vector $\pp\qq$.
The equilibrium of the joint system is again the equipartition,
$(\pp\qq)^{\rm eq}=\pp^{\rm eq}\qq^{\rm eq}$, that is the equilibrium is
multiplicative with respect to joining the systems if the latter are statistically 
independent.
The condition of additivity for the Lyapunov function (\ref{Lyap1})
of the joint system reads: 

\begin{equation}
\label{add}
H_h(\pp\qq)=H_h(\pp)+H_h(\qq).
\end{equation}

This functional equation has two special solutions, corresponding to the convex functions,
$h_1(x)=x\ln x$, and
$h_2(x)=-\ln x$. We denote $H_1=H_{x\ln x}$ and $H_2=H_{-\ln x}$, respectively. 
Whereas the function $H_1$ corresponds to the classical (additive) 
Boltzmann-Gibbs-Shannon entropy,
we shall demonstrate here the additivity of  $H_2$. Indeed,

\begin{eqnarray*}
H_2(\pp\qq)&=&-\sum_{\{ij\}=1}^{PQ}P^{-1}Q^{-1}\ln(PQp_iq_j)\\
&=&-\ln(PQ)-\sum_{i=1}^{P}\ln p_i-\sum_{j=1}^Q \ln q_j\\
&=&-\sum_{i=1}^{P}P^{-1}\ln(Pp_i)-\sum_{j=1}^QQ^{-1}\ln(Qq_j)\\
&=&H_2(\pp)+H_2(\qq).
\end{eqnarray*}

Neglecting the irrelevant constant and constant factors, and using Eq.\  (\ref{gen}),
we finally arrive at the one-parametric family of additive convex Lyapunov functions  
(\ref{Lyap1})
for master equation with $N$ states:
\begin{equation} 
\label{result}
H_{\alpha}=
(1-\alpha)\sum_{i=1}^N p_i\ln p_i-\alpha \frac{1}{N}\sum_{i=1}^N\ln p_i, \ 0\le\alpha\le 1
\end{equation}
The one-parametric family of additive Lyapunov functions is the central point of our further 
discussion.
Several remarks are in order:

{\it Remark 1} In the thermodynamic limit, which in the case considered here corresponds 
formally to 
$N\to\infty$, for any $\alpha$, we have $H_{\alpha}\to(1-\alpha)H_1$. That is,
the non-classical contribution due to $H_2$ becomes significant only if the system is 
not too close
to the thermodynamic limit. In the thermodynamic limit survives only the 
classical Boltzmann-Gibbs-Shannon
contribution.

{\it Remark 2} It is not difficult to prove that the family (\ref{result}) 
exhausts all the possible additive Lyapunov functions
of the form (\ref{Lyap1}) (up to adding a constant, and a multiplication with a constant factor): 
Indeed,
the classical treatment of the additivity condition requires averaging the vector  
function $\ln \pp$ which
can be done either using $\pp$ or $\pp^{\rm eq}$. 
The latter is the distinguished probability distribution
which is, same as $\pp$, multiplicative with respect to joining the statistically independent
subsystems. 
Relevance of master equation, and hence of kinetic rather than of static picture,
 to our derivation of the
one-parametric family (\ref{result}) is apparent: This enables to consider {\it two}
sets of probabilities, the ``current'' $\pp$, and the ``final'' $\pp^{\rm eq}$ (the equipartition here).

Other convex functions which are additive under joining statistically
independent systems do exist, for example, the R\'enyi entropy function \cite{Renyi70} but
they are not of the form (\ref{Lyap1}) (that is not of the so-called ``trace form'',
cf.\  Ref.\ \cite{Wehrl78}).  For this reason, such functions
fall out of our discussion since the proofs of the inequalities (\ref{production1}) and
(\ref{production2}) are not valid for them. 

{\it Remark 3} Function $H_2$ is not defined 
(and, consequently, any of the function $H_{\alpha}$, $\alpha\ne 0$ is not defined)
if one of the probabilities $p_i$ equals to zero.
The classical Boltzmann-Gibbs-Shannon solution to the additivity equation
is distinguished by the property of continuity at $p_i=0$.
This is a blueprint of the long-tail features  (see next section).
Work with the family of entropies (\ref{result})  assumes preserving
additivity on the expense of abandoning the continuity of the entropy functions 
on closed intervals $0\le p_i\le 1$, and its replacement  by continuity
on semi-open intervals, $0< p_i\le 1$.

{\it Remark 4} To the best of our knowledge, the entropy function 
\begin{equation}
\label{Burg}
S_2=-H_2=\sum_{i=1}^N\ln p_i
\end{equation}
has been first considered by Burg in the context of applications
of information theory to geophysical problems \cite{Burg67}, \cite{Burg72}.
Recently, the Burg entropy (\ref{Burg}) has been used to
construct illustrations of the  entropic lattice Boltzmann method \cite{KFOe99} 
 in Ref.\  \cite{Boghosian01}.
However, we failed to find a reference to the one-parametric family (\ref{result})
prior to Ref.\ \cite{Gorban84}. Whereas in Ref.\ \cite{Gorban84} the one-parametric
family (\ref{result}) has been mentioned as just the solution to the additivity condition,
its relevance to describing  effects of finiteness in statistical systems
has not been duly discussed.

{\it Remark 5} If the equilibrium $\pp^{\rm eq}$ of the Markov chain differs from 
the equipartition but is multiplicative under joining statistically independent subsystems, the 
one-parametric family (\ref{result}) generalizes to the following:

\begin{equation}
\label{result2}
H_{\alpha}=
(1-\alpha)\sum_{i=1}^N p_i\ln \left(\frac{p_i}{p_i^{\rm eq}}\right)-
\alpha\sum_{i=1}^N p_i^{\rm eq}\ln \left(\frac{p_i}{p_i^{\rm eq}}\right).
\end{equation}

\section{Longer tails: An example}\label{test}
In this section we want to explicitly work out an example in order to demonstrate
that the entropies of the family (\ref{result}) indeed describes the long tails for $\alpha\ne1$.
When the discrete system of states is addressed, as it is done here, the meaning of the
long tail has to be understood as broadening of  the distribution functions.

Without loss of generality, we shall work in this section with the one-parametric set,
\begin{equation}
\label{result1}
H_{\alpha}=
(1-\alpha)\sum_{i=1}^N p_i\ln p_i-\alpha \sum_{i=1}^N\ln p_i, \ 0\le\alpha\le 1.
\end{equation}
We shall consider first the microcanonic ensemble,
that is, the minimizer of $H_{\alpha}$ under the constraint of fixed normalization,
$\sum_{i=1}^Np_i=1$. It is straightforward to see that, for any admissible value of the
parameter $\alpha$, the microcanonic state is the equipartition, as expected.

In order to address the canonic ensemble, we introduce energies of the states $E_i\ge0$,
and find the minimum of $H_{\alpha}$ (\ref{result1}) under the constraints,
\begin{eqnarray}
\label{constraints}
\sum_{i=1}^Np_i&=&1,\label{norm}\\ 
\sum_{i=1}^{N}E_ip_i&=&U. \label{energy}
\end{eqnarray}
Denoting the solution $\pp^{(\alpha)}$, we find, for $\alpha\ne1$,
\begin{equation}
\label{soln}
p_i^{(\alpha)}\exp\left\{-\frac{\alpha}{(1-\alpha)p_i^*}\right\}=\exp\{\lambda-\beta E_i\},
\end{equation}
where $\lambda$ and $\beta$ are Lagrange multipliers, corresponding to the
constraints (\ref{constraints}). In order to address the effect of $\alpha\ne0$, we shall
restore to a perturbation theory around the Boltzmann-Gibbs-Shannon point $p_i^{(0)}$.
After a few  algebra, we get for $\alpha\ll1$, $\alpha>0$:

\begin{equation}
\label{can}
p_i^{(\alpha)}=p_i^{(0)}+\alpha\left(1-Np_i^{(0)}+(U-E_i)\frac{V-NU}{C-U^2}p_i^{(0)}\right),
\end{equation}
where
\begin{equation}
\label{canB}
p_i^{(0)}=\frac{1}{Z^{(0)}}e^{-\beta^{(0)}E_i},\ Z^{(0)}=\sum_{j=1}^Ne^{-\beta^{(0)}E_j},
\end{equation}
is the canonical distribution function for the Boltzmann-Gibbs-Shannon entropy
 (Lagrange multiplier
$\beta^{(0)}$ is expressed in terms of the average energy $U$ by the constraint (\ref{energy});
we do not need here the explicit expression $\beta^{(0)}(U)$), and
\begin{eqnarray}
V&=&\sum_{i=1}^NE_i,\label{V}\\
C&=&\sum_{i=1}^NE^2_ip_i^{(0)},\label{C}
\end{eqnarray}
that is,$V$ is the total energy of the states, and the 
denominator appearing in Eq.\ (\ref{can}), $C-U^2$, is the correlation of the energy levels
$E_i$ in the canonical state (\ref{canB}). We further denote,
\begin{equation}
B=\frac{V-NU}{C-U^2}.
\end{equation}
It can be argued that $B>0$: The total energy of the states, $V$, is not less
 (and in most of the relevant
cases, much larger) than the average energy $U$ times the number of states, whereas the
correlator $C-U^2$ is always positive.

Function (\ref{can}) is the first-order perturbation result, and it is not a positive
definite quantity. Yet, it is sufficient to our purpose here, since the question we want to address 
is as follows:
What is the sign  of the derivatives, $dp_i^{(\alpha)}/d\alpha\big|_{\alpha=0}$? Probabilities
of the canonical distribution (\ref{canB}) decay when $E_i$ exceed the average energy $U$, so,
by switching on the Burg component, do we see the ``raising'' of the populations
of this ``high-energy tail''? In order to see this, we obtain in  Eq.\ (\ref{soln}),
\begin{equation}
\frac{dp_i^{(\alpha)}}{d\alpha}\Big|_{\alpha=0}=A_ip_i^{(0)},
\end{equation}
where the factor $A_i$ is,
\begin{equation}
\label{factor}
A_i=Z^{(0)}e^{\beta^{(0)}E_i}-N-B(E_i-U).
\end{equation}
Factor $A_i$ amplifies populations of the states which 
are less populated in the standard canonical ensemble (\ref{canB}) if
$E_i$ satisfies the inequality, $E_i>\epsilon$, where $\epsilon$ is the solution to the equation:
\begin{equation}
\label{bc}
(1/Z^{(0)})e^{-\beta^{(0)}\epsilon}(N+B(\epsilon-U))=1.
\end{equation}

In order to make the situation even more transparent, we shall assume that the energies $E_i$ are
in a narrow band around the value $E>0$, that is $E_i=E+\delta_i$, $\sum_{i=1}^N\delta_i=0$, 
and $\delta_i\ll E$. 
All the quantities contributing to the expression (\ref{can}) can be then evaluated in terms
of expansion in $\delta_i$ (notice that the second-order perturbation in $\delta_i$ must be used
in order to compute the correlation $C-U^2$). We obtain, $B=\beta^{(0)}_E+o(\delta_i^2)$, and,
up to second order, eq.\ (\ref{bc}) reads
\begin{equation}
\label{bcnarrow}
\beta(N/2-1)\delta^2+\beta(1-N)\delta+\beta^2(N^{-1}-1/2)\sum_{i=1}^N\delta_i^2=0.
\end{equation}
For large $N$, this gives that factor (\ref{factor}) is larger than zero, and hence amplifies
 the populations
of the energy levels $E+\delta_i$, if
\begin{equation}
\label{broad}
\delta_i\ge 2/\beta^{(0)}_E.
\end{equation}
That is,  raising of the populations of 
higher energy levels is explicitly demonstrated in this example. We do not
discuss corrections for finite $N$ to the estimate (\ref{broad}) which are easily obtained from
Eq.\ (\ref{bcnarrow}).

Thus, we have demonstrated with explicit example that taking into account the Burg component 
in the one-parametric family (\ref{result}) indeed is able to describe  broadening of the 
canonical distribution
function. 
Appearance of the energy levels correlation in the above formula (\ref{can}) remarkably
resembles  recent results of application of Tsallis entropy to fitting 
experimental data in turbulence (see Ref.\ \cite{Beck02} and references cited therein).

Generalizations to quasi-equilibrium situation with more constraints is straightforward.
Also, if generalizations to a continuous case of states are addressed, the long tail feature of
the corresponding distribution functions becomes even more apparent. For example,
the counterpart of the Gaussian distribution function $P(x)$ 
(maximizer of the Boltzmann-Gibbs-Shannon entropy
under the constraint fixing the normalization and the variance) has the form,
$ P(x)\sim (\lambda+\beta x^2)^{-1}$. 
For it, algebraic decay at infinity precludes existence of the normalization,
and a cutoff is required.

To this end, we have argued that for systems out of the strict thermodynamic limit,
there exists a universal (that is, not explicitly dependent on details of interactions)
one-parametric family of additive entropy functions, which are able to describe,
at least in principle, the same long-tail effects as the Tsallis entropy.
In the remainder of this paper we shall discuss a different issue of how
additivity of the entropy  can {\it apparently} be violated if the description of the system is 
incomplete.

\section{Non-additivity  and incomplete description}\label{ghost}

Discussions  aimed at justifying a 
 non-additive dependence S(\pp) are sometimes conducted
in a rather obscure way: One argues that the entropy is non-additive
under joining the statistically independent subsystems because, {\it in reality},
these subsystems are not independent. 
Possible physical agents
which could lead to such a situation are occasionally mentioned, like long-range forces, 
for example, which seemingly justify statements like ``the concept of {\it independent}
subsystems does not make any sense, since all subsystems are interacting''  \cite{Beck02}.
However, to the best of our knowledge, 
a direct demonstration has never been done for any realistic
system.  Moreover, one should be more cautious in 
 rejecting  the ``concept of independent subsystems'',
especially when discrete systems as
above are  addressed, since this may 
lead  to a confrontation with the traditional axiomatic (Kolmogorov's) probability
theory which it is strongly based on this concept \cite{Kac57} (in the worst case, one should
abandon the concept of independent trials which is at the very hart of the definition
of probabilities). 

This all leads to a question: If the probability distribution over pairs of states $p_{ij}$ factors into 
products of distributions, $p_{ij}=q_ir_j$, but subsystems are dependent,
then where this dependence is hidden?
 What is this ``new'' notion of independence
which does not coincide with the classical, 
${\rm Prob} (A|B)={\rm Prob} (A)$ {\it means} $A$ is independent
of $B$? In order to answer this question, one should  realize  that such a situation of a ``hidden''
dependence, in fact,  has been  long known  in physics. This is the Pauli exclusion
principle. The corresponding Fermi-Dirac entropy has the well known form:

\begin{equation}
\label{Fermi}
S(\pp)=-\sum_i [p_i\ln p_i +(1-p_i)\ln (1-p_i)].
\end{equation}

This expression can be interpreted in the following way: With the electron gas, there
is associated a gas of ``places'' (holes). The state of the ensemble of this gas
of holes is uniquely determined by the ensemble of the electrons, 
$p_{i,{\rm hole}}=1-p_i$. If, for two subsystems of the electrons,
$p_{ij}=q_ir_j$, then, for the corresponding ensembles of holes, we have
$p_{ij,{\rm hole}}=1-q_ir_j$, and the corresponding product for the subsystems
reads,
\[ (1-q_i)(1-r_i)=1-q_i-r_j+q_ir_j\ne p_{ij,{\rm hole}}.\]
Therefore, subsystems of the electrons are dependent even for the multiplicative $p_{ij}=
q_ir_j$.

One can term this the effect of hidden components. It should be stressed that
we speak, in fact, about an incomplete description (both the ensembles are
uniquely related to each other), namely, that there are hidden components whose entropy
has to be taken into account. 

A different example, without any reference to quantum effects, is the entropy of monolayers
on the surface of a solid (see, e.\ g.\ \cite{Yablonskii91}). 
In the simplest case, the entropy density, up to constant factors and constants, has the form,

\begin{equation}
\label{monolayer}
S=-c_{AZ}\ln(c_{AZ}/c^*_{AZ})-c_{Z}\ln(c_Z/c^*_Z),
\end{equation}
where $A$ denotes molecules of the gas, $Z$ is the vacant position on the surface
(adsorbing center), $AZ$ is the adsorbed molecule, $c$ denotes corresponding
surface concentrations, $^*$ denotes equilibrium concentrations, and
since $c_Z+c_{AZ}={\rm const}$ (the number of places per unit area is conserved)
Eq.\ (\ref{monolayer})  is again  the Fermi-Dirac entropy, obtained without
any relation to quantum effects.

Thus, to sum up, the simplest known version of violating the additivity 
implies existence of subsystems of ``locations'', ``holes'', ``ghosts'' and like.
These subsystems occupy the same states as the ``observed'' 
system, with probabilities,

\begin{eqnarray}
q_i&=&1-ap_i,\ a\in[0,1],\ {\rm or}\nonumber\\
q_i&=&(1-a)+ap_i
\label{constraint}
\end{eqnarray}
(We have stressed two possible cases, with a positive and with a negative
constraint.) There might be several such hidden  subsystems, and thus
\begin{equation}
\label{mixture}
S=S(\pp)+\sum_{j}a_jS_j(\qq^{(j)}),
\end{equation}
where $j$ is the label of the hidden subsystem, and $a_j>0$. 
What the hidden subsystems could be? 
For example, they can describe various nonideal effects like excluded volume 
in various spaces (not obligatory in the physical $R^3$, as in the example
of adsorbing centers). Other interpretations are probably possible. Here we 
do not consider any specific examples. Rather, we want to emphasize remarkable 
approximation possibilities provided by expression (\ref{mixture})
when the entropies $S_{\alpha}$ are used. Indeed, already for just one hidden subsystem
we have four fitting parameters (two coefficients in front of the Burg component
for the system and for the ghost, one coefficient $a$ in the constraint (\ref{constraint}),
and $1-a$ for the ghost).
Because approximations to the experimental data obtained by the maximum entropy
principle under certain constraints, whereas the choice of these coefficients
is yet another optimization problem, it is not difficult, in principle, to organize
a procedure of choosing the parameters (learning or fitting) in such a way as
it is done in neural networks based on the error back propagation algorithm.
Each time, an intriguing question will be arising, as to how many ghosts are 
needed for a description with a given accuracy, and what is the physical interpretation
of those. 

\section{Hidden subsystems changing kinetics}\label{kinetics}

So far, all our considerations of the entropies, including either the Tsallis  entropy,
or the family $S_{\alpha}$, as well as of the entropy of the ghosts (\ref{mixture})
have left a nice option that they allow  to do nothing about  the master equation.
Namely, all these  entropies can be  used to describe all kinds
of incompletely known or restricted equilibria, or for constructing (generalized)
canonical ensembles of dynamically conserved or quasi-conserved quantities.
If the probability evolves in time according to the master equation, all these
entropy functions behave correctly, that is, they monotonically increase with the time.
This fact is well known, and it was reviewed above in section \ref{Markov}.

In other words, as long as the hidden subsystem is described by the same set of states, as 
the observed one,  no restrictions on the Markov kinetic equation arise.
However, if more freedom is allowed in the choice of the entropy,  kinetics has to be 
modified. Indeed,  for Markov chain satisfying the detail balance
condition (\ref{DB}), the natural condition which defines the equilibrium of the transition
$p_i\rightleftharpoons p_j$ can be written,

\begin{equation}
\label{equilibrium1}
\frac{\partial S}{\partial p_i}=\frac{\partial S}{\partial p_j}.
\end{equation}
For the entropy function of the form $S=-H_h$, where $H_h$ is given by equation (\ref{Lyap1}),
this writes,

\begin{equation}
\label{equilibrium2}
h'(p_i/p^{\rm eq}_i)=h'(p_j/p^{\rm eq}_j),
\end{equation}
 and, because of strict monotonicity, this results in the definition of the equilibrium,
$p_i/p_i^{\rm eq}=p_j/p_j^{\rm eq}$. Master equation (\ref{Markov1}) with the detail balance
condition can be rewritten in such a way as to make it apparently consistent with
the latter result: We denote,

\[ w_{ij}=q_{ij}p_j^{\rm eq},\]
then master equation (\ref{Markov1}) can be rewritten,
\begin{equation}
\label{Markov2}
\dot{p}_i=\sum_{j=1}^N w_{ij}[(p_j/p^{\rm eq}_j)-(p_i/p^{\rm eq}_i)].
\end{equation}

However, if the entropy of the hidden subsystem does not have the form
$S_h=-H_h$, with $H_h$ given by Eq.\ (\ref{Lyap1}), for example, if it
includes terms like,
\[ (a_0+\sum_{i=1}^Na_ip_i)\ln (a_0+\sum_{i=1}^Na_ip_i),\]
then condition (\ref{equilibrium1}) results in a more complicated equation, which, unlike
Eq.\ (\ref{equilibrium2}), brings in the dependence on all  the components of the vector $\pp$.
In that case, a model kinetic equation, more general than the master equations can be addressed.

There exist a universal way of constructing kinetic equations in a way consistent with the
given entropy. Let us introduce notation, $\mu_i=-\partial S/\partial p_i$, and let $\Psi(x)$ be
a monotonically increasing function. Then we define the rate of transitions $p_i\to p_j$ as
\begin{equation}
\label{MDDrate}
w_{ij}(\pp)\Psi(\mu_i),
\end{equation}
where $w_{ij}=w_{ji}$,  $w_{ij}\ge0$ is a symmetric matrix with nonnegative entries (which might
also be functions of the probability distribution $\pp$). Given the rates (\ref{MDDrate}),
kinetic equation takes the form:
\begin{equation}
\label{MDD}
\dot{p}_i=\sum_{j}w_{ij}(\pp)(\Psi(\mu_j)-\Psi(\mu_i)).
\end{equation}
Equation (\ref{MDD}) is a generalization of the  Marcelin -- De Donder kinetic formalism
(see, for instance,
\cite{Gorban84}, \cite{Yablonskii91}, \cite{Grmela81}, \cite{KGG02}).

Equation (\ref{MDD}) becomes especially natural to use if the entropy of the system has the form,
\begin{equation}
S=S_h+\tilde{S},
\end{equation}
where the part $S_h=-H_h$ has standard form (\ref{Lyap1}) for some convex function $h$ while
$\tilde{S}$ is the entropy of the (part of the)  hidden subsystem which does not have such form.
Then we put,
$\Psi(x)=[-h']^{-1}(x)$, that is, $\Psi$ is the inverse of the derivative $-h'$. 
With this, Eq.\ (\ref{MDD})
becomes:
\begin{equation}
\dot{p}_i=\sum_{j=1}^Nw_{ij}\left([-h']^{-1}(h'(p_j/p_j^{\rm eq})-
\partial\tilde{S}/\partial p_j)-[-h']^{-1}(h'(p_i/p_i^{\rm eq})-\partial\tilde{S}/\partial p_i)\right).
\label{MDDnew}
\end{equation}
This is the minimal extension of the master equation: If the hidden system can be 
described with the
same entropy ($\tilde{S}=0$), equation (\ref{MDDnew}) reduces to master equation.
However, in any case, extensions (\ref{MDD}) and (\ref{MDDnew}) are consistent with
the entropy increase in the kinetic processes.

\section{Discussion}\label{conclusion}

Once a classical statistical system is out of the thermodynamic limit, the exclusive character
of the Boltzmann-Gibbs-Shannon entropy  is lost, and classical ensembles are not equivalent
anymore.  Whereas  using the microcanonical ensemble for any
description of  finite systems may be most appropriate,
 this route is very complicated from a computational standpoint. 
 For that reason, seeking an entropic description of effects
of finiteness seems a relevant option. 

In this paper, we have demonstrated  that that there exists a unique one-parametric family
of entropy functions which are consistent  with the additivity of the entropy
under joining statistically independent subsystems. This family is essentially
the convex combination of the Boltzmann-Gibbs-Shannon entropy and
the Burg entropy. This family of entropy functions appears in a natural way as a distinguished
(by the additivity requirement) subset of the family of Lyapunov functions
of master equation. It has been demonstrated that the nontrivial contribution from
the  Burg component results in broadening of the high-energy tail of the 
canonical distribution function. The functional form of the deviation, and,
in particular, the appearance of the energy correlations indicates that
the maximum entropy approach successively used recently in the context of Tsallis entropy
may lead to similar results when the present entropy functions are used.
Detailed study of this option is left for the future work.

Finiteness of the classical statistical system is one option which calls for
non-classical entropies. A different (independent) option is the incompleteness
of the description. This has been demonstrated by analyzing
the classical example of Fermi-Dirac type of entropy, and a generalization
in the form of ``standard entropy for a multi-component mixture plus linear constraints''
has been suggested. 
Finally, we have suggested a minimal modification of master equation
consistent with the given entropy.

\bibliography{karlin}

\begin{thebibliography}{21}
\expandafter\ifx\csname natexlab\endcsname\relax\def\natexlab#1{#1}\fi
\expandafter\ifx\csname bibnamefont\endcsname\relax
  \def\bibnamefont#1{#1}\fi
\expandafter\ifx\csname bibfnamefont\endcsname\relax
  \def\bibfnamefont#1{#1}\fi
\expandafter\ifx\csname citenamefont\endcsname\relax
  \def\citenamefont#1{#1}\fi
\expandafter\ifx\csname url\endcsname\relax
  \def\url#1{\texttt{#1}}\fi
\expandafter\ifx\csname urlprefix\endcsname\relax\def\urlprefix{URL }\fi
\providecommand{\bibinfo}[2]{#2}
\providecommand{\eprint}[2][]{\url{#2}}

\bibitem[{\citenamefont{Tsallis}(1988)}]{Tsallis88}
\bibinfo{author}{\bibfnamefont{C.}~\bibnamefont{Tsallis}},
  \bibinfo{journal}{J.\ Stat.\ Phys.} \textbf{\bibinfo{volume}{52}},
  \bibinfo{pages}{479} (\bibinfo{year}{1988}).

\bibitem[{Tpa()}]{Tpage}
\eprint{http://tsallis.cat.cbpf.br/biblio.html}.

\bibitem[{\citenamefont{Vives and Planes}(2002)}]{Vives02}
\bibinfo{author}{\bibfnamefont{E.}~\bibnamefont{Vives}} \bibnamefont{and}
  \bibinfo{author}{\bibfnamefont{A.}~\bibnamefont{Planes}},
  \bibinfo{journal}{Phys.\ Rev.\ Lett.} \textbf{\bibinfo{volume}{88}},
  \bibinfo{pages}{020601} (\bibinfo{year}{2002}).

\bibitem[{\citenamefont{Karlin et~al.}(2002)\citenamefont{Karlin, Grmela, and
  Gorban}}]{KGG02}
\bibinfo{author}{\bibfnamefont{I.~V.} \bibnamefont{Karlin}},
  \bibinfo{author}{\bibfnamefont{M.}~\bibnamefont{Grmela}}, \bibnamefont{and}
  \bibinfo{author}{\bibfnamefont{A.~N.} \bibnamefont{Gorban}},
  \bibinfo{journal}{Phys. Rev. E} \textbf{\bibinfo{volume}{65}},
  \bibinfo{pages}{036128} (\bibinfo{year}{2002}).

\bibitem[{\citenamefont{Ruelle}(1969)}]{Ruelle}
\bibinfo{author}{\bibfnamefont{D.}~\bibnamefont{Ruelle}},
  \emph{\bibinfo{title}{Statistical Mechanics. Rigorous Results}}
  (\bibinfo{publisher}{Benjamin, New York}, \bibinfo{year}{1969}).

\bibitem[{\citenamefont{Koga}(1954)}]{Koga54}
\bibinfo{author}{\bibfnamefont{T.}~\bibnamefont{Koga}}, \bibinfo{journal}{J.\
  Chem.\ Phys.} \textbf{\bibinfo{volume}{22}}, \bibinfo{pages}{1633}
  (\bibinfo{year}{1954}).

\bibitem[{\citenamefont{Kogan}(1965)}]{Kogan65}
\bibinfo{author}{\bibfnamefont{A.~M.} \bibnamefont{Kogan}},
  \bibinfo{journal}{J.\ Appl.\ Math.\ Mech.} \textbf{\bibinfo{volume}{29}},
  \bibinfo{pages}{130} (\bibinfo{year}{1965}).

\bibitem[{\citenamefont{Lewis}(1967)}]{Lewis67}
\bibinfo{author}{\bibfnamefont{R.~M.} \bibnamefont{Lewis}},
  \bibinfo{journal}{J.\ Math.\ Phys.} \textbf{\bibinfo{volume}{8}},
  \bibinfo{pages}{1448} (\bibinfo{year}{1967}).

\bibitem[{\citenamefont{Gorban and Karlin}(1994)}]{GK94b}
\bibinfo{author}{\bibfnamefont{A.~N.} \bibnamefont{Gorban}} \bibnamefont{and}
  \bibinfo{author}{\bibfnamefont{I.~V.} \bibnamefont{Karlin}},
  \bibinfo{journal}{Physica A} \textbf{\bibinfo{volume}{206}},
  \bibinfo{pages}{401} (\bibinfo{year}{1994}).

\bibitem[{\citenamefont{Levermore}(1996)}]{Levermore96}
\bibinfo{author}{\bibfnamefont{C.~D.} \bibnamefont{Levermore}},
  \bibinfo{journal}{J.\ Stat.\ Phys.} \textbf{\bibinfo{volume}{83}},
  \bibinfo{pages}{1021} (\bibinfo{year}{1996}).

\bibitem[{\citenamefont{Gorban}(1984)}]{Gorban84}
\bibinfo{author}{\bibfnamefont{A.~N.} \bibnamefont{Gorban}},
  \emph{\bibinfo{title}{Equilibrium Encircling. Equations of Chemical Kinetics
  and Their Thermodynamic Analysis}} (\bibinfo{publisher}{Nauka, Novosibirsk},
  \bibinfo{year}{1984}).

\bibitem[{\citenamefont{R\'enyi}(1970)}]{Renyi70}
\bibinfo{author}{\bibfnamefont{A.}~\bibnamefont{R\'enyi}},
  \emph{\bibinfo{title}{Probability Theory}}
  (\bibinfo{publisher}{North-Holland, Amsterdam}, \bibinfo{year}{1970}).

\bibitem[{\citenamefont{Wehrl}(1978)}]{Wehrl78}
\bibinfo{author}{\bibfnamefont{A.}~\bibnamefont{Wehrl}}, \bibinfo{journal}{Rev.
  Mod. Phys.} \textbf{\bibinfo{volume}{50}}, \bibinfo{pages}{221}
  (\bibinfo{year}{1978}).

\bibitem[{\citenamefont{Burg}()}]{Burg67}
\bibinfo{author}{\bibfnamefont{J.~P.} \bibnamefont{Burg}}, \bibinfo{note}{paper
  presented at the 37th meeting of the Society of Exploration Geophysicists,
  Oklahoma City, 1967}.

\bibitem[{\citenamefont{Burg}(1972)}]{Burg72}
\bibinfo{author}{\bibfnamefont{J.~P.} \bibnamefont{Burg}},
  \bibinfo{journal}{Geophysics} \textbf{\bibinfo{volume}{37}},
  \bibinfo{pages}{375} (\bibinfo{year}{1972}).

\bibitem[{\citenamefont{Karlin et~al.}(1999)\citenamefont{Karlin, Ferrante, and
  \"Ottinger}}]{KFOe99}
\bibinfo{author}{\bibfnamefont{I.~V.} \bibnamefont{Karlin}},
  \bibinfo{author}{\bibfnamefont{A.}~\bibnamefont{Ferrante}}, \bibnamefont{and}
  \bibinfo{author}{\bibfnamefont{H.~C.} \bibnamefont{\"Ottinger}},
  \bibinfo{journal}{Europhys.\ Lett.} \textbf{\bibinfo{volume}{47}},
  \bibinfo{pages}{182} (\bibinfo{year}{1999}).

\bibitem[{\citenamefont{Boghosian et~al.}(2001)\citenamefont{Boghosian, Yepez,
  Coveney, and Wagner}}]{Boghosian01}
\bibinfo{author}{\bibfnamefont{B.~M.} \bibnamefont{Boghosian}},
  \bibinfo{author}{\bibfnamefont{J.}~\bibnamefont{Yepez}},
  \bibinfo{author}{\bibfnamefont{C.~P.} \bibnamefont{Coveney}},
  \bibnamefont{and} \bibinfo{author}{\bibfnamefont{A.}~\bibnamefont{Wagner}},
  \bibinfo{journal}{Proc.\ Roy.\ Soc.\ Lond. A MAT}
  \textbf{\bibinfo{volume}{457}}, \bibinfo{pages}{717} (\bibinfo{year}{2001}).

\bibitem[{\citenamefont{Beck}(2002)}]{Beck02}
\bibinfo{author}{\bibfnamefont{C.}~\bibnamefont{Beck}},
  \bibinfo{journal}{Europhys.\ Lett.} \textbf{\bibinfo{volume}{57}},
  \bibinfo{pages}{329} (\bibinfo{year}{2002}).

\bibitem[{\citenamefont{Kac}(1957)}]{Kac57}
\bibinfo{author}{\bibfnamefont{M.}~\bibnamefont{Kac}},
  \emph{\bibinfo{title}{Probability and Related Topics in Physical Sciences}}
  (\bibinfo{publisher}{Interscience, NY}, \bibinfo{year}{1957}).

\bibitem[{\citenamefont{Yablonskii et~al.}(1991)\citenamefont{Yablonskii,
  Gorban, Bykov, and Elokhin}}]{Yablonskii91}
\bibinfo{author}{\bibfnamefont{G.~S.} \bibnamefont{Yablonskii}},
  \bibinfo{author}{\bibfnamefont{A.~N.} \bibnamefont{Gorban}},
  \bibinfo{author}{\bibfnamefont{V.~I.} \bibnamefont{Bykov}}, \bibnamefont{and}
  \bibinfo{author}{\bibfnamefont{V.~I.} \bibnamefont{Elokhin}},
  \emph{\bibinfo{title}{Kinetic Models of Catalytic Reactions. Comprehensive
  Chemical Kinetics, V. 32, ed. by G G. Compton}}
  (\bibinfo{publisher}{Elsevier, Amsterdam}, \bibinfo{year}{1991}).

\bibitem[{\citenamefont{Grmela}(1981)}]{Grmela81}
\bibinfo{author}{\bibfnamefont{M.}~\bibnamefont{Grmela}},
  \bibinfo{journal}{Can.\ J.\ Phys.} \textbf{\bibinfo{volume}{59}},
  \bibinfo{pages}{698} (\bibinfo{year}{1981}).

\end{thebibliography}

\end{document}